\documentclass[prb,aps,twocolumn,floats,showpacs]{revtex4-2}
\usepackage{amsthm}     % 定理和证明环境
\usepackage{amsmath,amssymb,bm}
\usepackage{graphicx}
\usepackage{subcaption}
\usepackage{float}
\usepackage{makecell}
\usepackage{pifont}
 \usepackage{ragged2e}
 \usepackage{booktabs}   % 提供 \toprule, \midrule, \bottomrule
\usepackage{multirow}   % 提供 \multirow
\usepackage{xcolor}     % 提供 \textcolor
\usepackage[colorlinks=true, letterpaper=true, pdfstartview=FitV, linkcolor=blue, citecolor=blue, urlcolor=blue]{hyperref}

\usepackage{color}

\def\eqa{\begin{eqnarray}}
\def\eea{\end{eqnarray}}
\newcommand{\eq}{\begin{equation}}
\newcommand{\ee}{\end{equation}}

\usepackage{xeCJK}
\usepackage{physics}
\begin{document}
\title{Topological properties and phase diagram of the triangular Hofstadter model with staggered flux}
\author{Qi Gao$^{1}$}
\author{Wei Chen(陈薇)$^{1,2}$} \email{chenweiphy@nju.edu.cn}
\affiliation{$^{1}$National Laboratory of Solid State Microstructures and School of Physics, Nanjing University, Nanjing, China}
\affiliation{$^2$Collaborative Innovation Center of Advanced Microstructures, Nanjing University, Nanjing, China}

%\date{\today}
\begin{abstract}
We study the topological properties and phase diagram of the triangular Hofstadter model with staggered flux in adjacent triangles in this work. This lattice can be used to describe the low energy physics of the twisted bilayer transition metal dichalcogenides (TMD) in a certain range of the electric displacement field between the two layers. We show that the Hofstadter spectrum of this model is generally asymmetric except at specific staggered flux $3\phi=\frac{\pi}{2} \mod \pi$ due to 
an additional ${\cal P}$ symmetry at such $\phi$. Breaking the translation symmetry by dimerization lifts the ${\cal P}$ symmetry and results in rich topological phases in the system. The dimerized model with different rational external magnetic flux $\Phi_B=2\pi p/q$ has phase diagram with the following common features. For even $q$, the dimerized model generally has three gapped regimes. The one with small dimerization has finite Chern number and the other two have zero Chern number. For odd q, the model is gapped  with zero Chern number at any finite dimerization. For both $q$ even and odd, the two regimes with zero Chern number can be further characterized by the inversion symmetry of the parametrized one-dimensional chains of the system at $\phi=0 \mod \frac{\pi}{3}$, and one regime is topologically non-trivial and the other is trivial. Our results may be tested in twisted bilayer TMD with weak interaction or cold atom systems in optical lattice or photonic crystals achieved in recent experiments. 
\end{abstract}

\maketitle

\newpage

\section{Introduction}
Twisted two-dimensional (2D) moiré system, formed by rotating one single-layer 2D lattice relative to  the  other, has attracted significant attention in recent years due to their capability to host phenomena that are difficult to realize in conventional 2D lattices \cite{TBG_2011,TBG_2018_1,TBG_2018_2,TMD1,TMD2,TMD3}. One prominent example is the Hofstadter model with rational magnetic flux per unit cell \cite{Hofstadter_1976}. In ordinary 2D lattices, achieving a sizable flux per unit cell—without it being obscured by disorder—requires extremely high magnetic fields. In contrast, moiré superlattices possess unit cells that can be orders of magnitude larger, often by a factor of up to a thousand. As a result, appreciable rational magnetic flux can be realized under experimentally accessible magnetic fields, enabling the direct observation of Hofstadter physics in these systems \cite{Hofstadter_Experiment_1,Hofstadter_Experiment_2,Hofstadter_Experiment_3}.

A particularly interesting class of 2D moiré systems is twisted bilayer transition metal dichalcogenides (tTMD) \cite{TMD4,TMD5,TMD6}. Both numerical studies and experiments have demonstrated that in these systems, hopping amplitudes and interaction strengths are highly tunable, leading to a rich variety of emergent phases \cite{TMD7}. Of particular interest are two regimes controlled by the electric displacement field between the layers. One corresponds to a moiré lattice effectively described by a tight-binding model on a triangular lattice with staggered flux \cite{Stagger_Triangular_2023,C.D.Gong_2011,Tri_2026}. The other is described by a generalized Haldane model on a hexagonal lattice \cite{TMD1, TMD2}. These two regimes exhibit distinct topological characteristics and give rise to different phases under interaction.
While the Haldane model \cite{Haldane}  has been extensively studied as a paradigmatic example of a quantum anomalous Hall (QAH) system \cite{QAHE} and as a precursor to the fractional QAH effect in the presence of interactions at fractional filling, the triangular lattice with tunable staggered flux remains comparatively less explored. In this work, we focus on the latter case and demonstrate that, under a magnetic field with rational flux, this system hosts a rich variety of topological phases and properties, which may in turn give rise to diverse correlated states upon the inclusion of interactions.

\begin{figure}[t]  
   \includegraphics[width=0.475\textwidth]{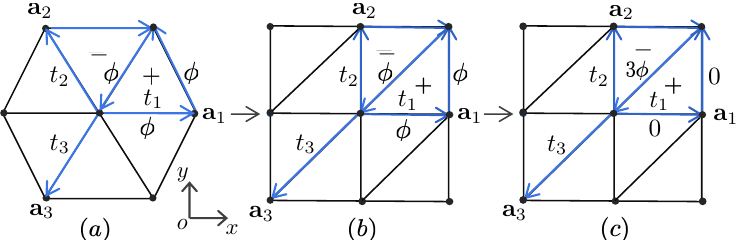} 
 \caption{  \justifying (a) Triangular lattice with staggered flux $\pm 3\phi$ in adjacent triangles; the hopping amplitudes along the three bonds of a triangle are respectively $t_1,t_2$ and $t_3$ with phase factor $\phi$. The lattice model in (a) is topologically equivalent to a square lattice model in (b) with nearest neighbor hopping plus next nearest neighbor hopping in one diagonal direction, which is again equivalent to the lattice model in (c).}
\label{triangular_lattice} 
\end{figure}

To investigate the properties of  tTMD in a magnetic field in the triangular lattice phase, we studied the triangular Hofstadter model with staggered flux $3\phi$ in adjacent triangles in this work, see Fig.\ref{triangular_lattice}(a). The triangular Hofstadter model with staggered flux can be mapped to a square lattice Hofstadter model with nearest neighbor (NN) hopping plus next nearest neighbor (NNN) hopping in one of the diagonal directions as shown in Fig.\ref{triangular_lattice}(b)\cite{Square(NNN)_Kohmoto_1990}. In contrast to the symmetric Hofstadter spectrum of the square lattice with only NN hopping, the Hofstadter spectrum of the triangular lattice with staggered flux $3\phi$ is asymmetric in general cases except at 
$\phi=\frac{\pi}{6}\mod \frac{\pi}{3}$. 
We show that this is due to a ${\cal P}$ symmetry of the model at these $\phi$ values. This symmetry results in $q$ gapless  Dirac points at $E=0$ for even $q$, where $q$ corresponds to the external magnetic flux $\Phi_B=2\pi p/q$. Breaking the translation symmetry, e.g., by dimerization of the hopping parameters in one direction, breaks the ${\cal P}$ symmetry and leads to rich topological phases in the system \cite{Dimerized_Square_2015}. 

We show that the dimerized triangular Hofstadter model has significantly different properties from its counterpart of square lattice with only NN hopping \cite{Xiao2024,Wen1989} due to the sub-lattice symmetry  breaking  in the former lattice. 
Combining a detailed study of the cases with $p/q=1/2, 1/3, 1/4$ and numerical analysis of more broad $p/q$ cases, we get the common feature of the phase diagram of the dimerized triangular Hofstadter model with staggered flux as follows. For even $q$, 
the dimerized model is gapped with zero Hall conductivity at dimerization $|\delta J|>|\delta J^\phi|$, where $\delta J$ describes  the difference of the hopping amplitudes  between two neighboring bonds in $\mathbf{a}_2$ direction and
 $\delta J^\phi$ is a $\phi$-dependent critical value corresponding to the band gap closing of the system. In the regime $|\delta J|<|\delta J^\phi|\neq 0$, the system is also gapped but with finite Chern number satisfying the constraint $C=\frac{q}{2}\mod q$. The two regimes $\delta J<-|\delta J^\phi|$ and $\delta J>|\delta J^\phi|$ with zero Hall conductivity may be distinguished by the topological invariant ${\cal N}$ due to the 
inversion symmetry \cite{Inversion_Symmetry} of parameterized one-dimensional (1D) chains at $k_1=-\frac{\pi p}{q}\mod \pi$ and $\phi=0 \mod \frac{\pi}{3}$, which reveals that one of the above two regimes is topologically non-trivial whereas the other is trivial. For odd $q$, there exist only two distinct regimes, i.e., $\delta J>0$ and $\delta J<0$, both of which have zero Hall conductivity, but one is topologically non-trivial and the other is topologically trivial.

The above results may be tested in tTMD in the weak interaction regime. By tuning the electric displacement field between the two layer, one may vary the staggered flux $3\phi$ and observe topological phase transitions with discontinuous Hall conductivity. The external magnetic field is another knob to tune the flux $\Phi_B$ which may also result in diverse topological phases. Other than the real 2D Moire materials, 
cold atom systems in optical lattice  \cite{Ultracold_Atomic, Cold_Atom, Ketterle2013} or photonic crystals \cite{Photonic_crystal} also provide tunable and controllable platform to observe the physics of the triangular Hofstadter model with staggered flux we study in this work. 

This paper is organized as follows. In Sec.\ref{Model}, we introduce the Hamiltonian of the triangular Hofstadter model with staggered flux, and analyze the symmetry of the Hofstadter spectrum of the model. In Sec.\ref{Dimerized_model}, we study the topological properties and phase diagram of the model with dimerization in one direction for the specific case with $p/q=1/2, 1/3, 1/4$. In Sec.\ref{Discussion}, we discuss the common feature of the dimerized model with general rational $p/q$ and the realization of the model in experiments. At last, we give a brief summary of this work.

\section{Model Hamiltonian and symmetry}\label{Model}
\subsection{Model Hamiltonian of the triangular Hofstadter model with staggered flux}

The moiré lattice of tTMDs in a certain range of the bilayer electric displacement field can be effectively described by a triangular lattice  shown in Fig.~\ref{triangular_lattice}(a), as demonstrated by the calculation of the electron local density of states of the system in \cite{TMD1}. Without the displacement field, the approximate inversion symmetry of  tTMD at small twist angel combined with the time reversal symmetry result in spin degeneracy of the electron bands at general $k$. Under a finite displacement field, the inversion symmetry of the system is broken and the upmost valence  bands of tTMD with spin-up and spin-down split, as shown in Fig.1(b) of Ref.\cite{HF_2021}. The spin splitting in the two valleys $K$ and $K'$ of tTMD is opposite due to the time reversal symmetry. The inversion symmetry breaking by the displacement field also leads to a finite phase of the hopping parameters in the system~\cite{TMD1, HF_2021}, which results in a staggered flux in adjacent triangles, as shown in Fig.~\ref{triangular_lattice}.
 Ignoring the inter-valley scatterings, the two valleys are independent. The effective model of the upmost valence band  of a single valley is then spin-polarized and can  be described by a spinless tight-binding model on a triangular lattice with staggered flux as \cite{Stagger_Triangular_2023}  
\begin{equation}
    H_{0}=-\sum\limits_{\boldsymbol{r}}\sum\limits_{n=1,2,3}t_{\mathbf{r}, \mathbf{r}+\boldsymbol{a}_n}e^{i\phi}c_{\boldsymbol{r}+\boldsymbol{a}_{n}}^{\dagger}c_{\boldsymbol{r}}+h.c.,
\end{equation}
where $\boldsymbol{a}_n, n=1,2,3$ are the nearest neighbor vectors labeled in Fig.\ref{triangular_lattice}(a), the hopping phase $\phi$ is tunable by the electric displacement field between the two layers as shown in \cite{TMD1} and results in staggered fluxes $\pm 3\phi$ crossing adjacent triangles as shown in Fig.~\ref{triangular_lattice}(a).

In this work we study the behavior of this model in a uniform external magnetic field with rational flux $\Phi_{B}=\frac{\sqrt{3}B}{2}\equiv \frac{2\pi p}{q}$ in each unit cell. The unit vectors of the lattice are $\boldsymbol{a}_1$ and $\boldsymbol{a_2}$ in Fig.\ref{triangular_lattice}(a), and the lattice sites $\mathbf{r}=m \boldsymbol{a}_1+n \boldsymbol{a}_2, m, n \in Z$. Here we have set the bond length of the triangular lattice $a=1$. We choose the gauge field $\boldsymbol{A}=\hat{y}(x+\frac{1}{\sqrt{3}}y){\mathbf B}$ for the magnetic field ${\mathbf B}$. The hopping parameters then get an extra Peierls phase in the magnetic field as 
$t_{\boldsymbol{r},\boldsymbol{r}^{\prime}}\to t_{\boldsymbol{r},\boldsymbol{r}^{\prime}}e^{i \int_{\boldsymbol{{r}}}^{\boldsymbol{r}^{\prime}}d\boldsymbol{x}\cdot \boldsymbol{A}(\boldsymbol{x})}$
and the corresponding Hofstadter model on the triangular lattice with staggered flux $\pm 3\phi$ in each triangle can be written as 
\begin{eqnarray}
     H^{\phi}&=&-\sum\limits_{m,n}(t_1e^{i\phi} c_{m+1,n}^{\dagger}+t_2e^{i\phi} c_{m,n+1}^{\dagger}e^{im\Phi_{B}} \nonumber\\
     && \     + t_3e^{i\phi}c_{m-1,n-1}^{\dagger}e^{-i\Phi_{B}(m-\frac{1}{2})})c_{m,n} +h.c..\label{eq:Tri_Hamiltonian}
\end{eqnarray}
This Hamiltonian exhibits a $4\pi$ periodicity in $\Phi_B$. It also describes the Hofstadter model on a square lattice with NN plus NNN hopping in a unique diagonal direction \cite{Square(NNN)_Kohmoto_1990}, with hopping phase $\phi$ on each step as shown in Fig.\ref{triangular_lattice}(b). The energy spectrum and Hall conductance of the case with $\phi=0$ has been studied in \cite{Square(NNN)_Kohmoto_1990}. In the following, we study the case with finite $\phi$.

By a gauge transformation  $c_{m,n}\to c_{m,n}e^{i(m+n)\phi},$ which is equivalent to a shift $(k_1,k_2)\to(k_1+\phi,k_2+\phi)$ in momentum space, followed by a dual transformation $t_1 \leftrightarrow t_2$, the Hamiltonian Eq.(\ref{eq:Tri_Hamiltonian}) becomes  
\begin{eqnarray}
 H^{\phi} &=& -\sum\limits_{m,n}(t_1  c_{m+1,n}^{\dagger}e^{in\Phi_{B}}+t_2 c_{m,n+1}^{\dagger} \nonumber\\
&&\   + t_3e^{3i\phi}c_{m-1,n-1}^{\dagger}e^{i\Phi_{B}(1/2-n)})c_{m,n} +h.c.,\label{squareH}
\end{eqnarray}
which describes the Hofstadter model on a square lattice with hopping phases shown in Fig.\ref{triangular_lattice}c and is invariant with $\phi\rightarrow \phi+2\pi/3$. At $t_3=0$, the Hamiltonian Eq.(\ref{squareH}) describes the square lattice Hofstadter model with NN hopping which has been studied extensively in previous works \cite{Hofstadter_1976,Zeromodes_Kohmoto_1989,Wen1989}. In this work, we focus on the case with $t_3\neq 0$ 
 and use the Hamiltonian Eq.(\ref{squareH}) in the following to describe the triangular Hofstadter model with staggered flux $\pm 3\phi$ per triangle.

For the rational flux $\Phi_B=2\pi p/q$, we choose a $1\times q$ magnetic unit cell (MUC). The momentum in the magnetic Brillouin zone (MBZ/BZ) can be written as $\boldsymbol{k}=k_1\boldsymbol{g}_1/2\pi +k_2\boldsymbol{g}_2/2\pi$ where $k_1\in [0,2\pi),k_2\in [0,\frac{2\pi}{q})$ and $\boldsymbol{g}_1,\ \boldsymbol{g}_{2}$ are the primitive reciprocal lattice vectors. The Hamiltonian matrix of Eq.(\ref{squareH}) after Fourier transformation to the momentum space of the MBZ is 
\begin{equation} \mathcal{H}^{\phi }(\boldsymbol{k})=\begin{pmatrix}
         A_1 & B_1^{*} &\cdots& \cdots &\cdots&B_{q}e^{iqk_2}\\
         B_1& A_2 & B_2^{*} &\cdots&\cdots &\cdots\\
         0 & B_2& A_3& B_3^{*}&\cdots &\cdots\\
         \cdots & \cdots& \cdots& \cdots&\cdots&\cdots \\
         0&\cdots& B_{q-3}& A_{q-2}&B^{*}_{q-2}&0\\
         0& 0  &\cdots &B_{q-2}&A_{q-1}& B_{q-1}^{*}\\
         B^{*}_{q}e^{-iqk_2}&\cdots&\cdots&\cdots&B_{q-1} &A_{q}
     \end{pmatrix}, \label{Hk}
\end{equation}
where
\begin{eqnarray}
&A_{j}=-2t_1\cos(k_1+2\pi j \frac{p}{q}),\ j=1,2,\cdots,q, \\
&B_{j}=-t_{2}-t_{3}e^{i(k_1+2\pi j \frac{p}{q}+\pi\frac{p}{q}-3\phi)},\ j=1,2,\cdots,q. \label{A_and_B} 
\end{eqnarray}

For any even $q$, we get 
\begin{equation}
{\rm Tr}[\mathcal{H}^{\phi }(\boldsymbol{k})]
%=-2 t_1\sum^q_{j=1}\cos{(k_1+2\pi j\frac{p}{q})}
=0. \label{eq:spectrum_sum}
\end{equation}
For $q=2$, we get immediately that the two energy bands of the system satisfy $E_1(k_1, k_2)=-E_2(k_1, k_2)$ for any $\phi$.
The energy spectrum $E$ at $\Phi_B=\pi$ is then symmetric with respect to $E=0$ as shown in Fig.\ref{Hofstadter_Spectrum}.

\subsection{Symmetry of the Hofstadter spectrum of the model}
\begin{figure}[htbp] 
\centering
\includegraphics[width=0.48\textwidth]{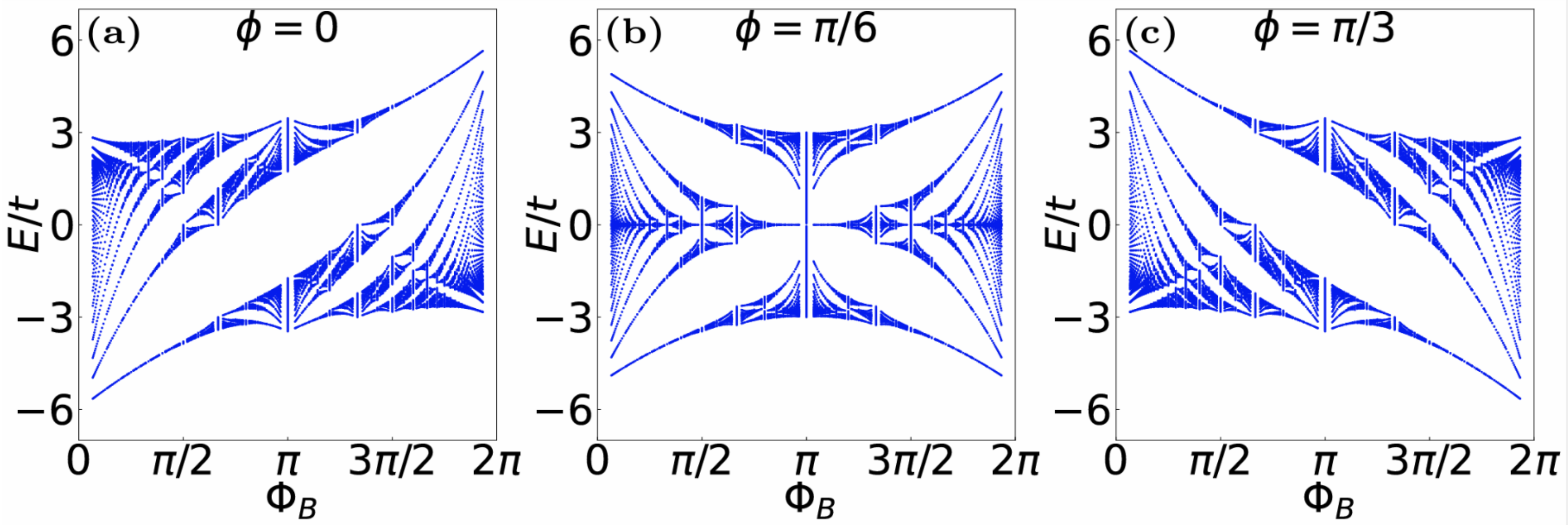} 
 \caption{  \justifying The Hofstadter spectrum of the triangular lattice with staggered flux $\phi$ at $t_1=t_2=t_3=t$~\cite{Code}.  (a)$ \phi=0$; (b)$\phi=\pi/6$; (c)$ \phi=\pi/3$. }
\label{Hofstadter_Spectrum}
\end{figure}

At $t_3=0$, the Hamitonian Eq.(\ref{squareH}) 
reduces to the square lattice Hofstadter model and  its Hofstadter spectrum is symmetric with respect to $E=0$ due to the sublattice symmetry \cite{Xiao2024,Wen1989}.
At $t_3 \neq 0$, the sublattice symmetry is lost and the Hofstadter spectrum of Eq.(\ref{squareH}) is generally asymmetric, 
%\textcolor{blue}{as for the case with $\phi=0$ in \cite{Square(NNN)_Kohmoto_1990} }
as shown in Fig.\ref{Hofstadter_Spectrum}(a) and (c).
However, we show in Fig.~\ref{Hofstadter_Spectrum}(b) that the Hofstadter spectrum of the system
becomes symmetric with respect to $E=0$ at $\phi=\pi/6+n \pi/3, n\in Z$ due to an additional symmetry \cite{Wen1989} in the Hamiltonian Eq.(\ref{squareH}).

 We fold the momentum in the original BZ to the MBZ and define a $q$ component spinor as $\Psi_{k_1,k_2}=
   (c_{k_1,k_2+\Phi_{B}}, c_{k_1,k_2+2\Phi_{B}}, \cdots, c_{k_1,k_2+q\Phi_{B}})^T$ where $(k_1, k_2)\in \rm MBZ$. As shown in Appendix A, the Hamiltonian Eq.(\ref{squareH}) after Fourier transformation to the momentum space can be written as 
   \begin{equation} H^{\phi}=\sum\limits_{k_1,k_2}\Psi_{k_1,k_2}^{\dagger}\mathcal{H}^{\phi}(k_1,k_2)\Psi_{k_1,k_2}, 
\end{equation}
 where
\begin{eqnarray}\label{H_translation}
&\mathcal{H}^{\phi}(k_1,k_2)= \nonumber\\
& -[t_1e^{ik_1}\mathbf{T}_1 +t_2e^{ik_2}\mathbf{T}_2+t_3e^{-i(k_1+k_2-3\phi)}\mathbf{T}_3]+h.c. \label{eq:H_trans}
\end{eqnarray}    
and $\mathbf{T}_1, \mathbf{T}_2, \mathbf{T}_3$ are the magnetic translation operations \cite{Zak} by $\mathbf{a}_1, \mathbf{a}_2, \mathbf{a}_3$ respectively, whose matrix representations are given in Appendix A.

At $\phi=\pi/6$, we rewrite Hamiltonian Eq.(\ref{eq:H_trans}) via a momentum shift  $(k_1,k_2)\to (q_1+\frac{\pi}{2},q_2+\frac{\pi}{2})$ with $(q_1,q_2)$ as the new momentum variables and get
 \begin{eqnarray} \label{eq:H_trans_2}
 & \mathcal{H}^{\frac{\pi}{6}}(q_1,q_2)=
 \notag\\ &-[i t_1e^{iq_1}\mathbf{T}_1+it_2e^{iq_2}\mathbf{T}_2-it_3e^{-i(q_1+q_2)}\mathbf{T}_3]+h.c..
 \end{eqnarray}
This Hamiltonian has 
a symmetry under the transformation $\mathcal{P}$ with matrix elements $\mathcal{P}_{ij}=\begin{cases}
    1, & i+j=0\mod q\\
    0, & \mathrm{otherwise}
\end{cases},\ i,j=1,2,\cdots q$. As shown in  Appendix A, 
\begin{equation}\label{P_T_commutation}
\mathcal{P}\mathcal{P}^\dag=\mathbf{I},\ \mathcal{P}\mathbf{T}_i\mathcal{P}^\dag=\mathbf{T}_i^{\dagger}, \ i=1,2,3.
\end{equation}
The Hamiltonian Eq.(\ref{eq:H_trans_2}) then satisfies
\begin{equation}
\mathcal{P}\mathcal{H}^{\frac{\pi}{6}}(q_1,q_2)\mathcal{P}^\dag=-\mathcal{H}^{\frac{\pi}{6}}(-q_1,-q_2). \label{Additional_Symmetry}
\end{equation}
This indicates that the eigenenergies $(E, q_1, q_2)$ and $(-E_1, -q_1, -q_2)$ appear in pairs and the Hofstadter spectrum is symmetric with respect to $E=0$ at $\phi=\pi/6$.
It is easy to check that at every $\phi=\frac{\pi}{6}+\frac{n \pi}{3}, n\in Z$, Eq.(\ref{Additional_Symmetry})
is satisfied so the Hofstadter spectrum is symmetric with a period of $\pi/3$ in $\phi$.

\subsection{Zero modes of the Hamiltonian at $\phi=\frac{\pi}{6}+\frac{n \pi}{3}$}
%The symmetry equation (\ref{Additional_Symmetry}) indicates there are zero-energy modes at $(q_1, q_2)=(0,0)$ for $\phi=\pi/6$.
 At $t_3=0$, the sublattice and translation symmetry in the square lattice Hofstadter model results in $q$ Dirac points at $E=0$ for even $q$ \cite{Wen1989,Zeromodes_Kohmoto_1989}. At $t_3 \neq 0$, the sublattice symmetry is broken. However, we show that the ${\cal P}$ symmetry at $\phi=\frac{\pi}{6}+\frac{n \pi}{3}, n\in Z$ can still result in Dirac points at $E=0$ for even $q$. Without loss of generality, we solve the zero modes of $\phi=\frac{\pi}{6}$ in the following.
 
 At $(q_1, q_2)=(0, 0)$, Eq.(\ref{Additional_Symmetry}) indicates that $\mathcal{H}^{\frac{\pi}{6}}(0,0)$ and $\mathcal{P}$ can have the same eigenstates. Assume $n_+ (n_-)$ are the number of zero energy eigenstates of $\mathcal{H}^{\frac{\pi}{6}}(0,0)$ with ${\cal P}=+1 ({\cal P}=-1)$,
the index~\cite{Wen1989}
\begin{equation}
I=n_+-n_-={\rm Tr}{\cal P}=\begin{cases}
    2, & q \in \rm  even,\\
    1, & q \in \rm  odd,
\end{cases}
\end{equation}
indicating that there are at least two (one) zero modes at $(q_1, q_2)=(0, 0)$ or $(k_1, k_2)=(\frac{\pi}{2}, \frac{\pi}{2})$ for $q$ even (odd).

For $q$ even, the two zero modes at $(q_1, q_2)=0$ form a Dirac point at $E=0$ with linear dispersion in general case. By the translation operation of $\mathbf{T}_2$, 
which shifts a Bloch state at $k=(k_1, k_2)$
to $k'=(k_1 + 2\pi \frac{p}{q}, k_2)$ for $\Phi_B=2\pi \frac{p}{q}$~\cite{constraints_2026}, one can get $q$ Dirac points at $E=0$ and momentum $(k_1, k_2)=((\frac{\pi}{2}+2\pi j \frac{p}{q})\mod 2\pi, \frac{\pi}{2}), j=0, 1, ..., q-1$.

The positions of the zero modes of $\mathcal{H}^{\frac{\pi}{6}}(k_1,k_2)$ can also be solved from the secular equation $\det([\mathcal{H}^{\phi}(\boldsymbol{k})]-E\mathbf{I}_{q\times q} )=0$. As shown in Appendix B, the secular equation for the zero modes of Hamiltonian Eq.(\ref{Hk}) is 
\begin{equation}
F(E=0)- f(k_1,k_2)=0,
\end{equation}
where 
\begin{eqnarray}
    f(k_1,k_2)&=& 2t_1^q\cos(qk_1)+2t_2^q\cos(qk_2) \nonumber\\
    &&- 2(-1)^{p+q}t_3^q\cos(qk_1+qk_2-3q\phi).
\end{eqnarray}
At $\phi=\frac{\pi}{6}+\frac{n\pi}{3}$,
\begin{eqnarray}
    F(E=0)&=& \begin{cases}
    2(-1)^{q/2}(t_1^{q}+t_2^{q}+t_3^{q}), & q \in \rm  even, \\
    0, & q \in \rm  odd.
    \end{cases}
    \label{fk2}
\end{eqnarray}
We then get the positions $(k_1,k_2)$ of the zero modes for $\phi=\frac{\pi}{6}+\frac{n\pi}{3}$  as follows:
\begin{align}
    (a)\ {\rm For}\ q&=4n, (k_1,k_2)=(0,0)\mod (\frac{2\pi}{q},\frac{2\pi}{q}), \\ 
    (b)\ {\rm For}\ q&=4n+2, (k_1,k_2)=(\frac{\pi}{q},\frac{\pi}{q})\mod (\frac{2\pi}{q},\frac{2\pi}{q}), \\ 
    (c)\ {\rm For}\ q&=2n+1, (k_1,k_2)=(\frac{\pi}{2q},\frac{\pi}{2q})\mod (\frac{\pi}{q},\frac{\pi}{q}). \label{Dirac_1}
\end{align}
All three cases include a solution at $(k_1, k_2)=(\frac{\pi}{2}, \frac{\pi}{2})$, as predicted by the index theorem. For case (a) and (b), i.e., $q$ even, the positions $(k_1, k_2)$ are the same as those obtained from the translation of $(\frac{\pi}{2}, \frac{\pi}{2})$ by $\mathbf{T}_2$. For $q$ odd, the solution from the secular equation
includes extra zero energy modes than those obtained from the translation symmetry. 
 At $q$ even, each zero mode corresponds to a Dirac point due to the degeneracy at $(k_1, k_2)$, whereas for odd $q$, the zero modes are non-degenerate at the corresponding momentum $(k_1, k_2)$.

If the translation symmetry $\mathbf{T}_i, i=1, 2, 3$  of the system is broken, the Hamiltonian at $\phi=\pi/6+n\pi/3, n\in Z$ can no longer be written as Eq.(\ref{eq:H_trans_2}) and ${\cal P}$ is no longer a symmetry operator. The Dirac points at $E=0$ for even $q$ are lifted and a gap opens at $E=0$. This gap opening results in interesting topological phases we study in the following.

\section{Topological properties of the dimerized model}\label{Dimerized_model}
The dimerization breaks the translation symmetry of the system and may arise from the formation of the charge density wave or strain in the system\cite{Stagger_Triangular_2023,HF_2021}. For simplicity, we dimerize the lattice along the $\boldsymbol{a}_{2}$ direction by introducing  alternating hopping amplitudes $J_1$ and $J_2$ along this direction. Due to the duality between $\boldsymbol{a}_1$ and $\boldsymbol{a_2}$ direction, the dimerization along $\boldsymbol{a_1}$ results in analog results.

For even $q$, the dimerization in $\boldsymbol{a}_2$ direction does not enlarge the MUC. For odd q, the MUC needs to be doubled to be periodic \cite{Dimerized_Square_2015}. 
The Hamiltonian matrix of the dimerized model in the momentum space then becomes an $N\times N$ matrix with 
\begin{equation}\label{dimension_N}
N = \begin{cases}
q, & q\geq 2\ \text{and\ even},\\
2q, & q>2\ \text{and\ odd},
\end{cases}
\end{equation}
and can still be written as Eq.(\ref{Hk}) but with $t_2$ in $B_j, j=1, 2, ..., N$ replaced by
\begin{equation}\label{eq:dimerized_hopping}
t_2=
\begin{cases}
J_{1}\equiv 1+\delta J, & j\in \mathrm{odd},\\ 
J_2\equiv 1-\delta J,& j\in \mathrm{even},
\end{cases}, j=1, 2, ..., N,
\end{equation}
where we have set $t_2\equiv 1$ before dimerization.

The Hamiltonian in momentum space can be viewed as a collection of dimerized 1D chains in $\boldsymbol{a}_2$ direction parameterized by the momentum $k_1$ \cite{Dimerized_Square_2015, Z2_2020}. At $t_3=0$, the above dimerized model reduces to the square lattice Hofstadter model and has been studied in \cite{Hofstadter_1976, Zeromodes_Kohmoto_1989,Wen1989, Dimerized_Square_2015}. 
At half filling, the Chern number of the gapped phases of this dimerized model is zero \cite{Dimerized_Square_2015, Xiao2024}. 
However,  a subset of parameterized 1D chains at specific $k_1$ values of this model preserves inversion symmetry at general rational flux $\Phi_B= 2\pi p/q$\cite{Dimerized_Square_2015}.
This inversion symmetry guarantees the presence of topologically protected end modes at dimerization $\delta J \equiv (J_1-J_2)/2 < \delta J_c<0$, where $\delta J_c$ is a critical value corresponding to the phase transition from a topologically non-trivial phase to a trivial phase.

For the triangular Hofstadter model with $t_3\neq 0$ and $\phi \neq 0$, the subset of inversion symmetric 1D chains at specific $k_1$ values does not exist for all $\phi$. The $k_1$ parameterized 1D Hamiltonian with inversion symmetry in $\mathbf{a}_2$ direction should satisfy the following relationship
 \begin{equation}\label{inversion_condition}
P\,\mathcal{H}^\phi(k_1,k_2)\,P^\dagger=\mathcal{H}^\phi(k_1,-k_2),
 \end{equation}
where $P$ is the inversion operator of the 1D chain, which is an anti-diogonal $N\times N$ matrix given by
$P_{ij}=\delta_{i, N+1-j}$.
By solving Eq.(\ref{inversion_condition}), we obtained the subset of 1D chains with inversion symmetry for the triangular Hofstadter model at
\begin{equation}
k_1=-\frac{\pi p}{q}\mod \pi \ {\rm and} \ \phi=\frac{n\pi}{3}, n \in Z.
\end{equation}

The topological properties of the 1D chains at these specific $k_1$ and $\phi$ can then 
be characterized by the topological invariant  ${\cal N}=|n_1-n_2|$, where $n_1$ and $n_2$ are the number of negative parities of the occupied states at the high symmetry momenta $k_2=0$ and $\pi/N$ respectively, with $N$ given in Eq.(\ref{dimension_N})\cite{Inversion_Symmetry}. At such high symmetry momenta $k_2$, $H(k_1, k_2)=H(k_1, -k_2)$ so the parity operator $P$ commutes with the Hamiltonian $H(k_1, k_2)$, i.e., $[P, H(k_1, k_2)]=0$ and the parity is well-defined at these momenta. For simplicity, we only study the dimerized model at half filling in this work. The properties of the system at other fillings can be analyzed in the same way.

Besides the above inversion symmetry at specific $k_1$ and $\phi$, we show 
in the following that the dimerized triangular Hofstadter model also possesses other symmetries and topological invariants depending on the flux $\Phi_B=2\pi p/q$, which results in more diverse topological phases in the system than the square lattice Hofstadter model. We demonstrate this by studying the specific cases of $p/q=\frac{1}{2}, \frac{1}{3}$ and $\frac{1}{4}$ in details in the following.

\subsection{Topological properties and phase diagram in the case of $p/q=1/2$}

\begin{figure}[htbp] \includegraphics[width=0.47\textwidth]{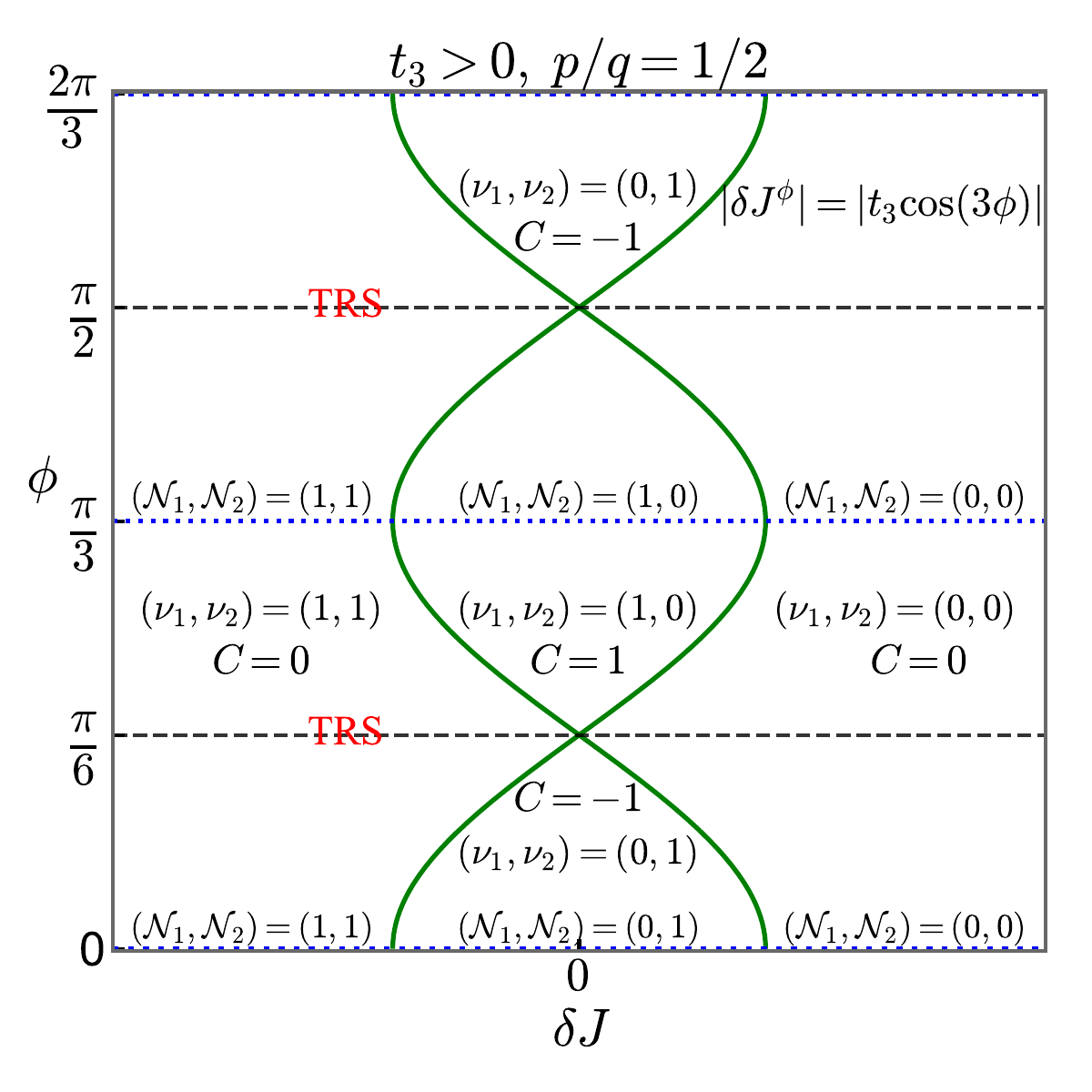} 
 \caption{\justifying  Phase diagram of the dimerized model at half filling for $p/q=1/2$ and $t_3>0$ over a period of $\phi$. The two curves defined by $|\delta J|=|t_3 \cos{3\phi}|$ correspond to the gap-closing boudaries. The system is gapped away from the two curves with the labeled Chern number $C$. The indices $(\nu_1, \nu_2)$ label the winding numbers at $k_1=\frac{\pi}{2}$ and $\frac{3\pi}{2}$. The indices $({\cal N}_1, {\cal N}_2)$ label the 
 topological invariant ${\cal N}$ at $k_1=\frac{\pi}{2}$ and $\frac{3\pi}{2}$ respectively for $\phi=0$ and $\frac{\pi}{3}$.
 The black dashed lines at $\phi=\frac{\pi}{6}$ and $\frac{\pi}{2}$ correspond to the time reversal symmetric case.
 The phase diagram of $t_3<0$ corresponds to the above diagram with $\phi\to \phi+\frac{\pi}{3}$. }
\label{phase_diagram_1}
\end{figure}

We first study the case with $p/q=1/2$. The Hamiltonian matrix of the dimerized model is 
\begin{equation}\label{dimerized_H} 
\mathcal{H}^{\phi}(k_1,k_2)=-\begin{pmatrix}
    -2t_1\cos{k_1} & B^*_1+B_2 e^{i2k_2} \\ B_1+B^*_2 e^{-i2 k_2} &2t_1\cos{k_1}
    \end{pmatrix}
\end{equation}
where $B_1$ and $B_2$ are defined in Eq.(\ref{A_and_B}) with dimerized $t_2$ in Eq.(\ref{eq:dimerized_hopping}). The Hamiltonian results in two bands. From the determinant of $\mathcal{H}^{\phi}(k_1,k_2)$, we get that at the condition $|2\delta J|\equiv|J_1-J_2|=|2 t_3 \cos{3\phi}|$, the gap of the two bands is closed. Otherwise, the gap is finite. The Chern number $C$ of the gapped phase at half-filling  is finite at $|J_1-J_2|<|2 t_3 \cos{3\phi}|$ and zero at  $|J_1-J_2|>|2 t_3 \cos{3\phi}|$, as shown in Fig.\ref{phase_diagram_1}.

\begin{figure}[tbp] 
\includegraphics[width=0.47\textwidth]{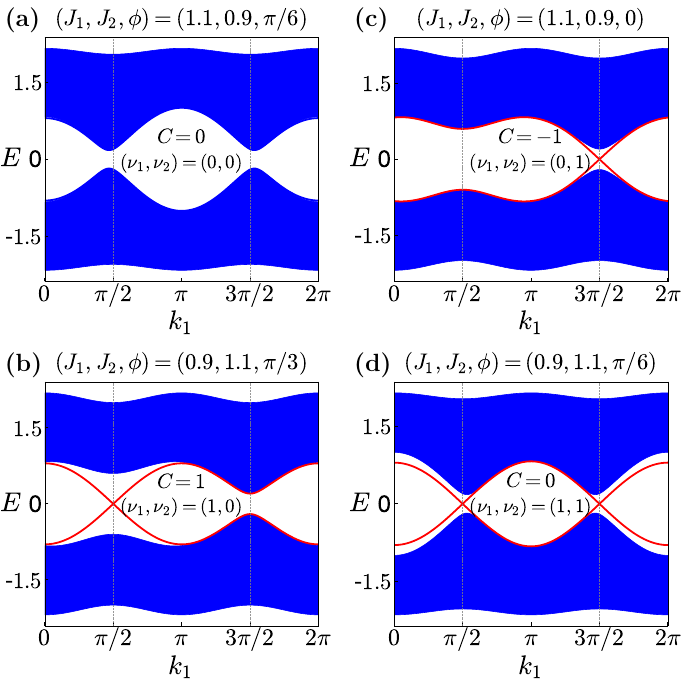}  
 \caption{\justifying  Band structure of the dimerized model with $p/q=1/2$ in a ribbon geometry of width $2N_2 a$ and $t_1=0.4,,t_3=0.2$ \cite{Code}. (a) Band structure in the regime  $J_1-J_2>|2 t_3 \cos{3\phi}|$. (b) and (c) Band structure in the regime $|J_2-J_1|<|2 t_3 \cos{3\phi}|$. (d)  Band structure in the regime $J_1-J_2<-|2 t_3 \cos{3\phi}|$. }
\label{edge_states}
\end{figure}

We plotted the energy bands of the dimerized model with $p/q=1/2$ in a ribbon geometry of width $2N_2 a, N_2\in Z$ in the $\mathbf{a}_2$ direction in the different regimes as a function of $k_1$, as shown in Fig.\ref{edge_states}. 
 In the regime $|J_1-J_2|<|2 t_3 \cos{3\phi}|$, there exists a pair of edge states in the half-filling gap due to the finite Chern number in this regime, as shown in Fig.\ref{edge_states}(b) and (c).
However, at $J_1-J_2<-|2 t_3 \cos{3\phi}|$, though the Chern number of the dimerized system is zero, there still exists a pair of robust edge modes pinned at zero energy and momentum $k_1=\frac{\pi}{2}$ and $\frac{3\pi}{2}$, as shown in Fig.\ref{edge_states}(d). This indicates that the system at half-filling in this regime is  topologically non-trivial.

 In the case $t_3=0$, i.e., in the dimerized square lattice Hofstadter model, the topological states at $C=0$ and $\delta J<\delta J_c<0$ can be characterized by the inversion symmetry of the 1D chains at $k_1=\frac{\pi}{2}$ and $k_1=\frac{3\pi}{2}$ for $p/q=1/2$. Specifically, a pair of edge states will be pinned at $k_1=\frac{\pi}{2}$ or (and) $k_1=\frac{3\pi}{2}$ when the topological invariant ${\cal N}$ at such $k_1$ is finite.
  However, at $t_3\neq 0$, this inversion symmetry exists only for $\phi=\frac{n\pi}{3}, n\in Z$ and the corresponding topological invariant ${\cal N}_1, {\cal N}_2$ at $k_1=\frac{\pi}{2}$ and $\frac{3\pi}{2}$ are labeled in Fig.\ref{phase_diagram_1}
  for $\phi=0$ and $\frac{\pi}{3}$. In Fig.\ref{edge_states}(b) and (c), a pair of edge states are pinned at $k_1$ for which the topological invariant ${\cal N}$ is finite in the case $\phi=0$ and $\frac{\pi}{3}$.

  Yet from Fig.\ref{edge_states}(d), we observe pinned edge states at $k_1=\frac{\pi}{2}, \frac{3\pi}{2}$ in the half filling gap even for $\phi\neq n\pi/3$. 
 In the following, we show that at $p/q=1/2$ and $t_3 \neq 0$, the dimerized Hamiltonian Eq.(\ref{dimerized_H}) 
 has a chiral symmetry at $k_1=\frac{\pi}{2}$ and $k_1=\frac{3\pi}{2}$ for any $\phi$ which 
 can be used to characterize the topological properties of the system in the whole parameter space.

At $k_1=\frac{\pi}{2}$ and $\frac{3\pi}{2}$, the diagonal terms of Hamiltonian Eq.(\ref{dimerized_H}) vanish and the 1D chains parameterized by $k_1$ at these values posses a chiral symmetry satisfying $\{\sigma_{3},\mathcal{H}^{\phi}(k_1,k_2)\}=0$. The Hamiltonian at $k_1=\frac{\pi}{2}$ and $\frac{3\pi}{2}$ can be written as 
\begin{equation}
\mathcal{H}^{\phi}(k_1,k_2)
    =-[h_{1}(k_2)\sigma_{1}+h_{2}(k_2)\sigma_{2}]\equiv -\boldsymbol{h}(k_2)\cdot \boldsymbol{\sigma},
\end{equation}
where $\sigma_{1,2}$ are the Pauli matrices and $h_1(k_2)=J_1+J_2\cos(2k_2)\pm 2t_3\sin(k_2)\sin(k_2-3\phi),\ h_2(k_2)=-J_2\sin(2k_2)\pm 2t_3\cos(k_2)\sin(k_2-3\phi)$ for $k_1=\frac{\pi}{2}$ and $\frac{3\pi}{2}$ respectively. 
 For the chiral symmetric 1D chains at $k_1=\frac{\pi}{2}$ and $\frac{3\pi}{2}$, there exists a topological invariant for the gapped phase, i.e., the winding number \cite{Winding_Number}
\begin{align} 
    \nu &=\frac{1}{2\pi}\oint \frac{h_1\partial_{k_2}h_2-h_2\partial_{k_2}h_1}{h_1^2+h_2^2}d k_2,
\end{align}
which counts the number of $\boldsymbol{h}(k_2)$ encircling the origin over one period of $k_2$.

We computed the winding number $\nu_1$ and $\nu_2$ at $k_1=\frac{\pi}{2}$ and $\frac{3\pi}{2}$ 
respectively for the gapped phases with $C=0$ and $\pm 1$, as shown in Fig.\ref{phase_diagram_1}. The winding number $(\nu_1, \nu_2)$ reveals that only in the regime $J_1-J_2>|2 t_3 \cos{3\phi}|$ where both $C=0$ and $(\nu_1, \nu_2)=(0, 0)$, the system at half filling is topologically trivial and there are no edge states in the gap as shown in Fig.\ref{edge_states}(a). In all the other gapped regimes, one of $\nu_1$ and $\nu_2$ or both of them are nonzero. The system in these regimes is topologically non-trivial with the crossing of a pair of edge bands pinned at the chiral symmetric points $k_1=\frac{\pi}{2}$ or (and) $\frac{3\pi}{2}$ with non-zero winding number, as shown in Fig.\ref{edge_states}(b)-(d). Crossing the boundary lines $|J_1-J_2|=|2 t_3 \cos{3\phi}|$, the gap of the system closes and reopens, and a topological phase transition takes place.

Since the dimerized Hamiltonian Eq.(\ref{dimerized_H}) with $p/q=1/2$ still satisfy ${\rm Tr} [\mathcal{H}^{\phi}(k_1,k_2)]=0$, its spectrum satisfies $E_1(k_1, k_2)=-E_2(k_1, k_2)$ for any $\phi$, i.e., the spectrum is symmetric with respect to $E=0$. The degenerate crossing points of the edge bands then must locate at $E=0$.

The two dashed lines locating at $\phi=\pi/6$ and $\phi=\pi/2$ in Fig.\ref{phase_diagram_1} correspond to the time reversal symmetric case for $p/q=1/2$. For the reason, the Chern number of the gapped system along these two lines is always zero. Crossing the gap-closing point $J_1=J_2$, a topological phase transition happens with changing winding number though the Chern number remains zero. 

\subsection{Topological properties and phase diagram in the case of $p/q=1/4$}

For $p/q=1/4$, the dimerized Hamiltonian takes the form
\begin{widetext}
  \begin{equation}\label{Hamiltonian_4}
    \mathcal{H}_{4\times 4}^{\phi}(k_1,k_2)=-\begin{pmatrix}
    2t_1\cos(k_1+\frac{\pi}{2}) & J_1+t_3e^{-i(k_1+\frac{3\pi}{4}-3\phi)} & 0 & (J_2+t_3e^{i(k_1+\frac{\pi}{4}-3\phi)})e^{4ik_2}\\
    J_1+t_3e^{i(k_1+\frac{3\pi}{4}-3\phi)} & 2t_1\cos(k_1+\pi)&J_2+t_3e^{-i(k_1+\frac{5\pi}{4}-3\phi)}& 0\\
    0& J_2+t_3e^{i(k_1+\frac{5\pi}{4}-3\phi)}& 2t_1\cos(k_1+\frac{3\pi}{2})& J_1+t_3e^{-i(k_1+\frac{7\pi}{4}-3\phi)}\\
    (J_2+t_3e^{-i(k_1+\frac{\pi}{4}-3\phi)})e^{-4ik_2}&0&J_1+t_3e^{i(k_1+\frac{7\pi}{4}-3\phi)}&2t_1\cos(k_1) \end{pmatrix}.
    \end{equation}
\end{widetext}

The chiral symmetry of the dimerized Hamiltonian for $p/q=1/2$ at specific $k_1$ values is lost for $p/q=1/4$. However, the  $k_1$ parameterized 1D Hamiltonian still has inversion symmetry at $k_1=-\frac{\pi}{4}$ and $\frac{3\pi}{4}$ for $\phi=n\pi/3, n\in Z $. The topological properties of the system at $\phi=n \pi/3$ can then be characterized by this inversion symmetry.
In a period of  $\phi$, we only need to consider the case of $\phi=0$ and $\frac{\pi}{3}$.

\begin{table}[t]
    \centering
    \caption{\justifying Eigenenergy and parity of the eigenstates at $k_1=-\frac{\pi}{4}$ and $k_2=0, \frac{\pi}{4}$ for $\phi=0$ and $p/q=1/4$.}
    \resizebox{\columnwidth}{!}{%
        \begin{tabular}{c|c|c}
            \toprule
            $(k_1,k_2)$  &  Eigenenergy & Parity \\
            \midrule
            \multirow{4}{*}{$\left(-\frac{\pi}{4},0\right) $}
            & $1-\delta J-\sqrt{(1+\delta J)^2+(t_1-\sqrt{2}t_3)^2+t_1^2}\equiv E_{11}$ & $-$ \\
            & $1-\delta J+\sqrt{(1+\delta J)^2+(t_1-\sqrt{2}t_3)^2+t_1^2}\equiv E_{12}$ & $-$ \\
            & $-1+\delta J-\sqrt{(1+\delta J)^2+(t_1+\sqrt{2}t_3)^2+t_1^2}\equiv E_{13}$ & $+$ \\
            & $-1+\delta J+\sqrt{(1+\delta J)^2+(t_1+\sqrt{2}t_3)^2+t_1^2}\equiv E_{14}$ & $+$ \\
            \midrule
            \multirow{4}{*}{$\left(-\frac{\pi}{4}, \frac{\pi}{4}\right) $}
            & $-t_3-\sqrt{(\sqrt{2}\delta J-t_1)^2+(t_1+\sqrt{2})^2+t_3^2}\equiv E_{21}$ & $-$ \\
            & $-t_3+\sqrt{(\sqrt{2}\delta J-t_1)^2+(t_1+\sqrt{2})^2+t_3^2}\equiv E_{22}$ & $-$ \\
            & $t_3-\sqrt{(\sqrt{2}\delta J+t_1)^2+(t_1-\sqrt{2})^2+t_3^2}\equiv E_{23}$  & $+$ \\
            & $t_3+\sqrt{(\sqrt{2}\delta J+t_1)^2+(t_1-\sqrt{2})^2+t_3^2}\equiv E_{24}$  & $+$ \\
            \bottomrule
        \end{tabular}%
    }    \label{parity}
\end{table}

\begin{figure}[t] 
\centering
\includegraphics[width=0.48\textwidth]{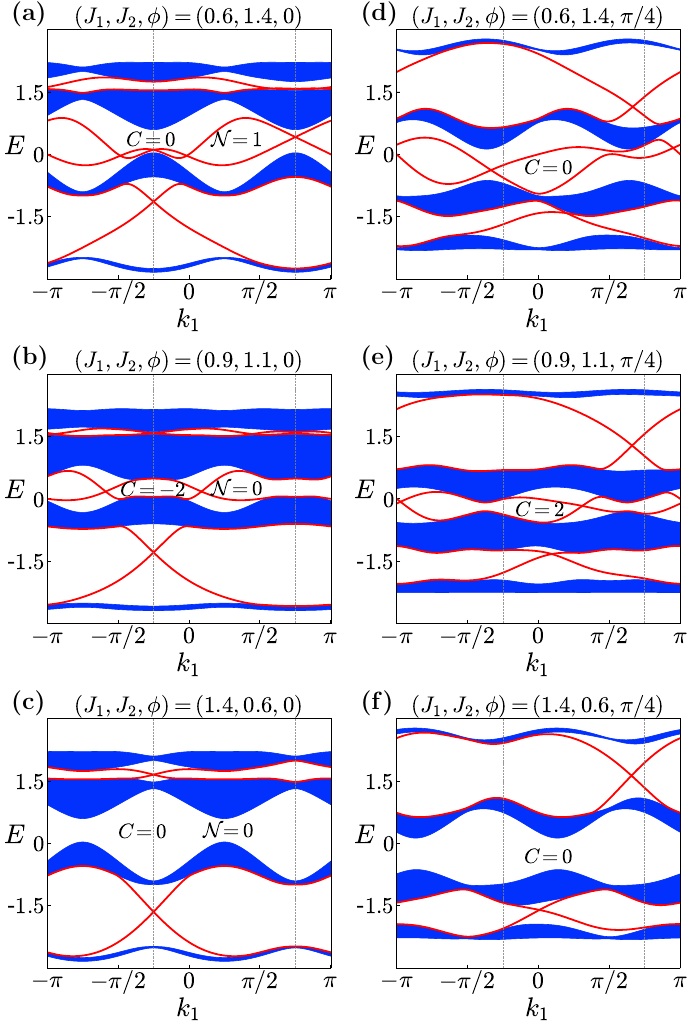} 
 \caption{\justifying The band structure of the dimerizied model with $p/q=1/4$ in a ribbon geometry of width $4N_2 a$ and $t_1= t_3=0.5$ \cite{Code}. (a) and (d) Band structure in the regime $\delta J<-|\delta J_c \cos{3\phi}|$ for $\phi=0$ and $\frac{\pi}{4}$ respectively, where $\delta J_c=-0.2293$ for $t_1=t_3=0.5$. (b) and (e) Band structure in the regime $|\delta J|<|\delta J_c \cos{3\phi}|$  for $\phi=0$ and $\frac{\pi}{4}$ respectively. (c) and (f) Band structure in the regime $\delta J>|\delta J_c \cos{3\phi}|$ for $\phi=0$ and $\frac{\pi}{4}$ respectively. }
\label{q=4_band}
\end{figure}

Without loss of generality, we show the calculation of the topological invariant ${\cal N}$ at $k_1=-\frac{\pi}{4}$ and $\phi=0$ at half filling in the following.
 For the 1D Hamiltonian with specific $k_1$, the high symmetry point in the momentum space with well-defined parity locates at $k_2=0$ and $\frac{\pi}{4}$ for $p/q=1/4$.
 The eigenenergies and parities of the eigenstates at $k_2=0$ and $\frac{\pi}{4}$ are shown in Table \ref{parity} respectively for $k_1=-\frac{\pi}{4}$ and $\phi=0$. For general $t_1, t_3>0$,
\begin{equation}
  E_{21}<0,\ E_{22}>0,\ E_{23}\le 0,\ E_{24}>0
\end{equation} 
in Table \ref{parity} and the number of negative parity eigenstates at $k_2=\frac{\pi}{4}$ at half filling is $n_2=1$. For the eigenstates at $k_2=0$, denoting $c\equiv 1-\delta J$, $ R_{1}\equiv \sqrt{(2-c)^2+(t_1-\sqrt{2}t_3)^2+t_1^2},\ R_{2}\equiv \sqrt{(2-c)^2+(t_1+\sqrt{2}t_3)^2+t_1^2}$ and comparing the eigenenergies of the four eigenstates, we get the number of negative parity states at $k_2=0$ at half filling as 
\begin{equation}
     n_1=\begin{cases}
          1, \ c<(R_1+R_2)/2,\\
          0, \ c>(R_1+R_2)/2.
     \end{cases}
\end{equation}
The topological invariant ${\cal N}$ at $k_1=-\frac{\pi}{4}$ and $\phi=0$ is then 
\begin{equation}\label{top_inv_N}
    \mathcal{N}=\vert n_1-n_2\vert=\begin{cases}
        0, \ c<(R_1+R_2)/2,\\
        1, \ c>(R_1+R_2)/2.
    \end{cases}
\end{equation}

By calculation, the topological invariants ${\cal N}$ at $k_1=-\frac{\pi}{4}$ and $\frac{3\pi}{4}$ for $\phi=0$ and $\frac{\pi}{3}$ are all the same as in Eq.(\ref{top_inv_N}). For $\phi=0$ and $\frac{\pi}{3}$, we then get at $c>(R_1+R_2)/2$, the system is topologically non-trivial with ${\cal N}=1$ at half filling with a pair of edge states pinned at $k_1=-\frac{\pi}{4}$ and $\frac{3\pi}{4}$, as shown in Fig.\ref{q=4_band}(a). At $c<(R_1+R_2)/2$, the topological invariant ${\cal N}=0$ at half filling and there are no edge states pinned at $k_1=-\frac{\pi}{4}$ and $\frac{3\pi}{4}$, as shown in Fig.\ref{q=4_band}(b) and (c). A topological phase transition with the change of ${\cal N}$ occurs at 
$c=(R_1+R_2)/2$, i.e.,
\begin{equation}\label{critical_delta_J}
2(1-\delta J)=\sqrt{(1+\delta J)^2+a}+\sqrt{(1+\delta J)^2+b},
\end{equation}
where $a\equiv (t_1-\sqrt{2}t_3)^2+t_1^2$, $b\equiv (t_1+\sqrt{2}t_3)^2+t_1^2$.

At $t_3=0$, the phase transition occurs at $\delta J_c=-t_1^2/2$ and the system is topologically non-trivial (trivial) at $\delta J<-t_1^2/2\ (\delta J> -t_1^2/2)$, which is consistent with the result for the dimerized square lattice Hofstadter model studied in \cite{Dimerized_Square_2015}. At $t_3 \neq 0$, the critical value  $\delta J_c$ at phase transition can be solved from Eq.(\ref{critical_delta_J}). At small $\delta J \ll 1$, we get 
\begin{equation}
\delta J_c \approx \frac{2-(\sqrt{1+a}+\sqrt{1+b})}{2+\frac{1}{\sqrt{1+a}}+\frac{1}{\sqrt{1+b}}}.
\end{equation}
At $\delta J<\delta J_c$, ${\cal N}=1$ and at $\delta J>\delta J_c$, ${\cal N}=0$.
Since Eq.(\ref{critical_delta_J}) remains the same by exchanging $t_1$ and $t_3$, the dependence of $\delta J_c$ on $t_1$ and $t_3$ is the same. 

\begin{figure}[htbp] 
\centering
\includegraphics[width=0.48\textwidth]{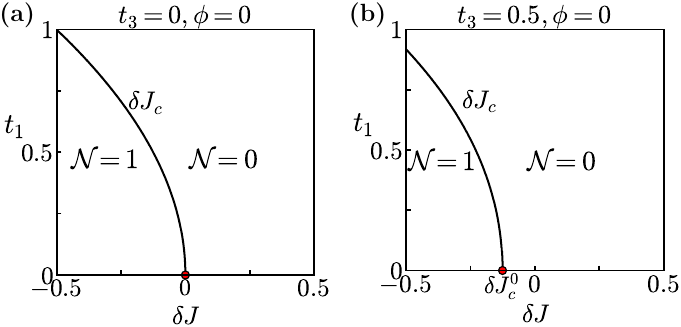} 
 \caption{\justifying The phase diagram of the topological invariant ${\cal N}$ for $p/q=1/4$ and $\phi=0$. The curve of $\delta J_c$ is solved numerically from Eq.(\ref{critical_delta_J}) and corresponds to the phase boundary from ${\cal N}=0$ to ${\cal N}=1$. (a)The phase diagram with fixed $t_3=0$, for which $\delta J_c=-t_1^2/2$. This corresponds to the case of the dimerized square lattice Hofstadter model with only NN hopping. (b)The phase diagram with fixed $t_3=0.5$. The critical value $\delta J_c^0=-t^2_3/2$ at $t_1=0$. The phase diagram for $\phi=\pi/3$ is the same as for $\phi=0$. }
\label{q=4_phase_diagram}
\end{figure}

In Fig.\ref{q=4_phase_diagram}, we plotted the phase diagram in terms of the topological invariant ${\cal N}$ for the case of $p/q=1/4$ and $\phi=0$, where the phase boundary $\delta J_c$ is solved from Eq.(\ref{critical_delta_J}) numerically. The phase diagram with $t_3=0$ in Fig.\ref{q=4_phase_diagram}(a) is the same as that for the square lattice Hofstadter model in Ref.\cite{Dimerized_Square_2015} with $\delta J_c=-t_1^2/2$.
At $t_3 \neq 0$, the phase diagram is modified as shown in Fig.\ref{q=4_phase_diagram}(b). The critical value of $\delta J_c$ at the phase transition point at $t_1=0$ is $\delta J^0_c=-t^2_3/2$.
The phase diagram for $\phi=\frac{\pi}{3}$ is the same as for $\phi=0$.

\begin{figure}[htbp] 
\centering
\includegraphics[width=0.45\textwidth]{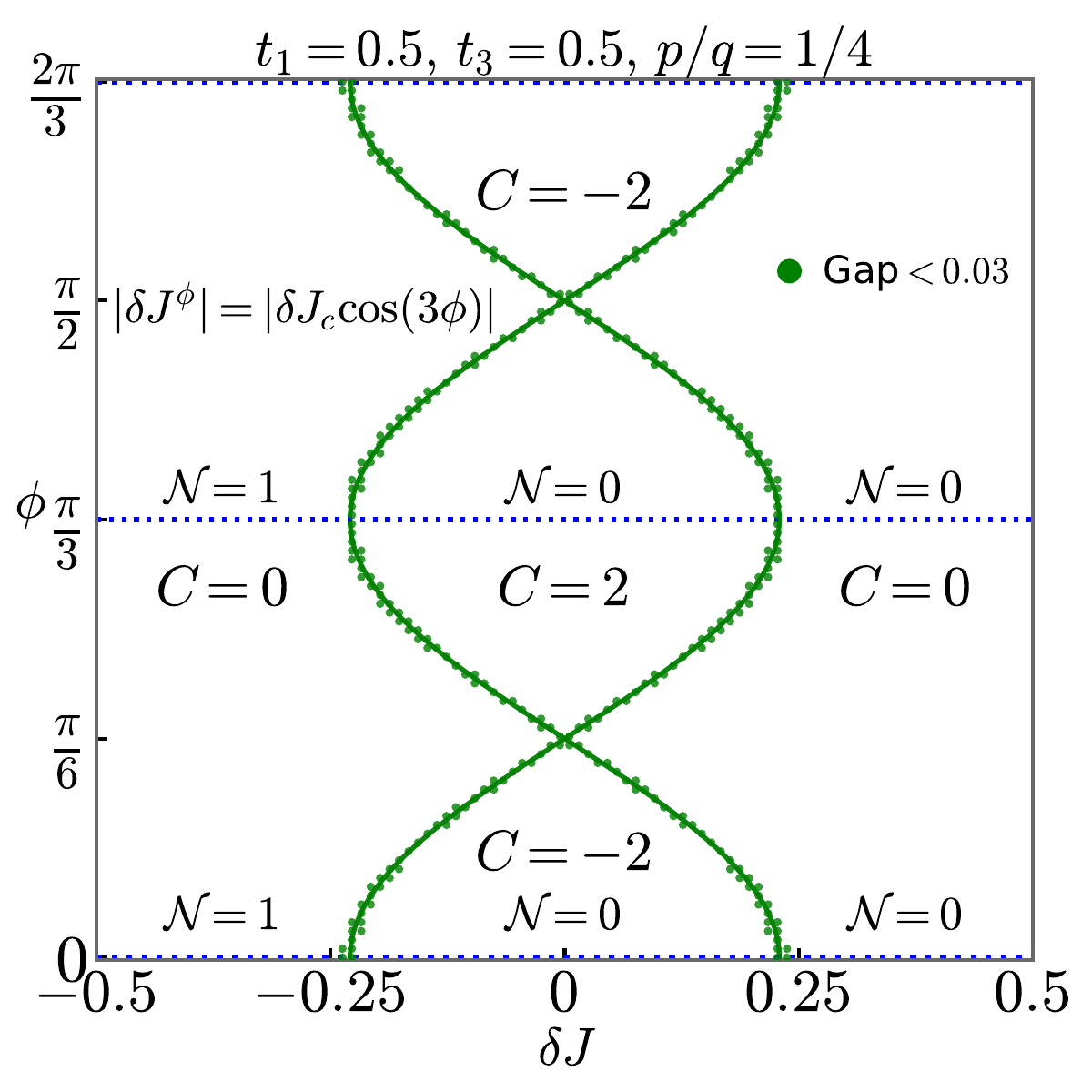} 
 \caption{\justifying The phase diagram of the dimerized model for $p/q=1/4$ with fixed $t_1, t_3 \neq 0$ and varying $\delta J$ and $\phi$ ($t_2
 \equiv 1$).\cite{Code} The two green curves $|\delta J|=|\delta J_c \cos{3\phi}|$ are fittings to the gap closing points obtained numerically, where $\delta J_c=-0.2293$ for $t_1=t_3=0.5$ and equals to $\delta J_c$ obtained from Eq.(\ref{critical_delta_J}). Here $C$ labels the Chern number in different gapped regimes and ${\cal N}$ labels the topological invariant due to the inversion symmetry of the parameterized 1D chains at $\phi=0 \mod \frac{\pi}{3}$.  }
\label{q=4_Chern number}
\end{figure}

For general $\phi$, to study the topological properties of the dimerized model at half filling, we computed the Chern number of the system numerically because it is hard to calculate it analytically for $p/q=1/4$.

For given nonzero $t_1, t_3$ and varying $\delta J$ and $\phi$, the gap closing points at half filling fit two curves $|\delta J|=|\delta J_c \cos{3\phi}|$ very well, as shown in Fig.\ref{q=4_Chern number}, where $\delta J_c$ is solved from Eq.(\ref{critical_delta_J}).
 %For the special case with $t_3=0$, the Chern number is zero in all the gapped phases due to the sublattice symmetry in this case \cite{Xiao2024}. In the following, we focus on the case with $t_3\neq 0$. 
 In the regime $|\delta J|<|\delta J_c \cos{3\phi}|$, the Chern number is finite, i.e., $C=2$ or $-2$. At $|\delta J|>|\delta J_c \cos{3\phi}|$, the Chern number is zero. 
From Fig.\ref{q=4_Chern number}, we can see that at $\phi=n \pi/3, n\in Z$, the gap closing at $\delta J=- |\delta J_c|$
accompanies both the change of the Chern number $C$ and the topological invariant ${\cal N}$ whereas the gap closing at $\delta J=|\delta J_c|$ accompanies only the change of the Chern number $C$. Though the Chern number $C=0$ in both  regimes $\delta J<-|\delta J_c \cos{3\phi}|$ and $\delta J>|\delta J_c \cos{3\phi}|$ in Fig.\ref{q=4_Chern number}, the topological properties of the two regimes are different due to different ${\cal N}$ at $\phi=0, \frac{\pi}{3}$. In the regime $\delta J<-|\delta J_c \cos{3\phi}|$, ${\cal N}=1$ for the 1D chains  at half filling at $k_1=-\frac{\pi}{4}$ and $\frac{3\pi}{4}$ and $\phi=0, \pi/3$  so the system is topologically non-trivial with edge states pinned at such $k_1$ values at $\phi=0, \frac{\pi}{3}$, as shown in Fig.\ref{q=4_band}(a). Without gap closing during the continuous variation of $\phi$ and $\delta J$ in this regime, the system remains topologically non-trivial, which can be verified by the existence of robust edge states in this regime shown in Fig.\ref{q=4_band}(d) though the edge states are no longer pinned at $k_1=-\frac{\pi}{4}, \frac{3\pi}{4}$. In contrast, in the regime $\delta J>|\delta J_c \cos{3\phi}|$, both $C=0$ and ${\cal N}=0$ at $\phi=\frac{n\pi}{3}$ at half filling, and the system is topologically trivial without edge states in the half filling gap for all $\phi$, as shown in Fig.\ref{q=4_band}(c) and (f). As a consequence, though the Hall conductance in both regimes of $\delta J<-|\delta J_c \cos{3\phi}|$ and $\delta J>|\delta J_c \cos{3\phi}|$ is zero, the longitudinal conductance at half filling  of the former is finite whereas of the latter is zero. In the regime $|\delta J|<|\delta J_c \cos{3\phi}|$, the Chern number $C$ is finite at half filling but ${\cal N}=0$ so the system is topologically non-trivial with edge states at the half filling gap but they are not pinned at $k_1=-\frac{\pi}{4}, \frac{3\pi}{4}$, as shown in Fig.\ref{q=4_band}(b) and (e).

\subsection{Topological properties and phase diagram of the dimerized model with odd $q$}\label{odd_q}
When $q$ is odd, the magnetic unit cell is doubled after the dimerization, which results in $2q$ bands in the system. If we artificially set the dimerization to zero, the $2q$ bands merge to $q$ bands and at half filling, there is no gap. For any finite dimerization, the gap is open at half filling. Numerical results show that the Chern number is zero in the gapped phase for both $\delta J>0$ and  $\delta J<0$. For simplicity, we only present the properties of the case $p/q=1/3$ in details in the following. 
\begin{figure}[htbp] 
\centering
\includegraphics[width=0.45\textwidth]{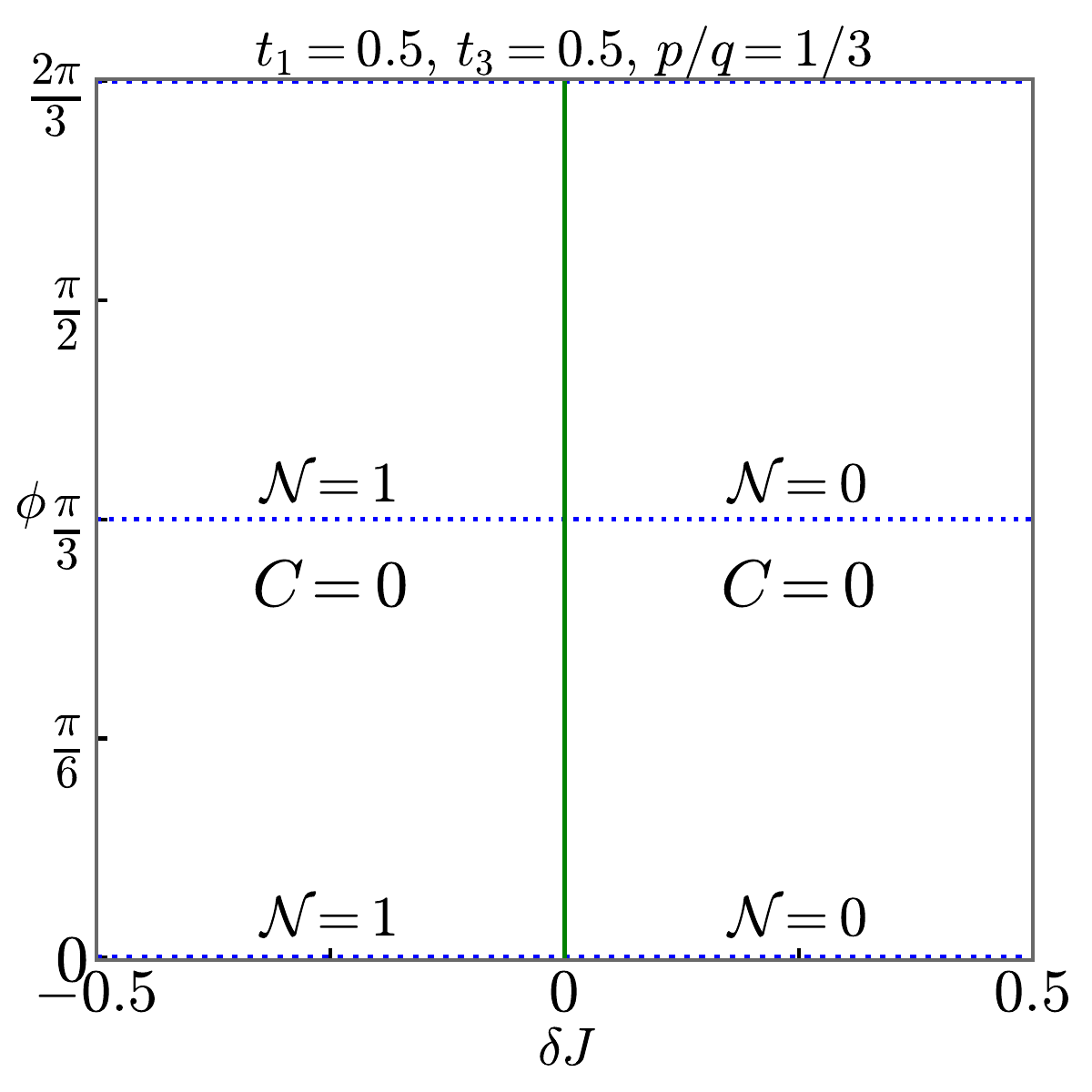} 
 \caption{\justifying The phase diagram of the dimerized model for $p/q=1/3$ with fixed $t_1, t_3$ and varying $\delta J$ and $\phi$ ($t_2
 \equiv 1$). The green line at $\delta J=0$ labels the gap closing points. In both gapped regimes $\delta J>0$ and $\delta J<0$, the Chern number $C=0$. In the regime $\delta J<0$, the topological invariant ${\cal N}$ due to inversion symmetry of the parameterized 1D chains at $\phi=0$ and $\frac{\pi}{3}$ is $1$ and it is zero in the regime $\delta J>0$. }
\label{q=3_Chern_number}
\end{figure}

As presented above, the 1D dimerized chain parameterized at $k_1=-\frac{\pi}{3}\mod \pi,\ \phi= \frac{n\pi}{3}, n\in Z$ has inversion symmetry for $p/q=1/3$. The topological properties of the dimerized model at $\phi=n \pi/3$ can then be characterized by the topological invariant ${\cal N}$ at $k_1=-\frac{\pi}{3}\mod \pi$ defined in the previous subsection. We get that for the half filling case, at $\delta J<0$, ${\cal N}=1$ and at $\delta J>0$, ${\cal N}=0$ at every 
$k_1=-\frac{\pi}{3}\mod \pi$ for all $\phi= \frac{n\pi}{3}$. The Chern number $C$ and topological invariant ${\cal N}$ for the $p/q=1/3$ case at half filling are shown in Fig.\ref{q=3_Chern_number}.

\begin{figure}[htbp] 
\centering
\includegraphics[width=0.48\textwidth]{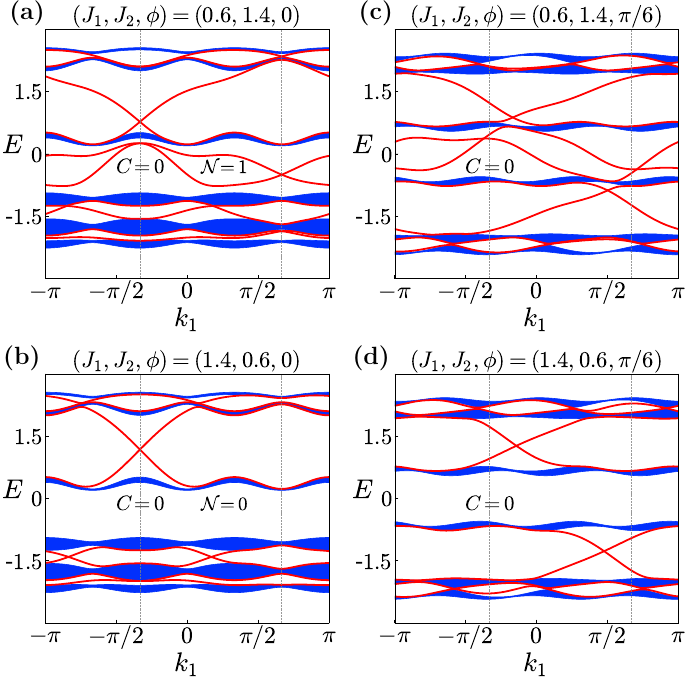} 
 \caption{\justifying The  band structure  of the dimerized model with $p/q=1/3$ in a ribbon geometry of width $6N_2 a$ and $t_1=t_3=0.5$ \cite{Code}.(a)and (c)Band structure in the regime $\delta J<0$ for $\phi=0$ and $\frac{\pi}{6}$ respectively. (b) and (d)Band structure in the regime $\delta J>0$ for $\phi=0$ and $\frac{\pi}{6}$ respectively.}
\label{q=3_band}
\end{figure}

For $\phi=\frac{n \pi}{3}, n\in Z$, at $\delta J<0$ a pair of edge states at half filling are then pinned at $k_1=-\frac{\pi}{3}\mod \pi$ due to the finite ${\cal N}$ as shown in Fig.\ref{q=3_band}(a), whereas at $\delta J>0$, there are no edge states at half filling
  as shown in Fig.\ref{q=3_band}(b) because ${\cal N}=0$. 
For $\phi \neq \frac{n\pi}{3}, n\in Z$, the inversion symmetry of the parameterized 1D chain is lost and ${\cal N}$ is not well-defined. However, in the regime $\delta J <0$ ($\delta J>0$),
the system remains topologically non-trivial (trivial) without gap closing with the continuous variation of $\phi$ and $\delta J$. This can be verified by the energy bands in the two regimes at $\phi \neq \frac{n \pi}{3}, n\in Z$. As shown in Fig.\ref{q=3_band}(c) and (d), for $\phi=\pi/6$, there are a pair of edge states at half filling at $\delta J<0$ indicating topological non-triviality of the state whereas there are no edge states at half filling at $\delta J>0$ indicating topological triviality in this case. Since ${\cal N}$ is not well-defined at $\phi \neq  \frac{n \pi}{3}$, the edge states at half filling at $\delta J<0$ in this case is not pinned at  $k_1 =-\frac{\pi}{3}\mod \pi$.

Numerical calculation reveals that the topological properties of the dimerized model for general odd $q$ are similar to the case of $p/q=1/3$. At  half filling, the system is  gapless at $\delta J=0$ and gapped at $\delta J \neq 0$. The 1D chains parameterized at $k_1=-\pi p/q \mod \pi$ and $\phi=n \pi/3, n\in Z$ have inversion symmetry with topological invariant ${\cal N}=1$ at $\delta J<0$ and ${\cal N}=0$ at $\delta J >0$.
 For the reason, the gapped phase at $\delta J<0$ is topologically non-trivial with edge states at the half filling gap and the gapped phase at $\delta J>0$ is topologically trivial at half filling without edge states in the gap.

\section{Discussions}\label{Discussion}
We studied the dimerized triangular Hofstadter model with other external rational flux numerically. The case for general odd $q$ has been presented in the Section \ref{odd_q}. For the reason,  we discuss the general case  with even $q$ in the following.

We first note that the case with $t_3=0$ is significantly different from the case with $t_3\neq 0$ due to the sublattice symmetry at $t_3=0$. This sublattice symmetry results in $C=0$ at half-filling for all the gapped phases of the dimerized model with  even $q$ as proved in Ref.\cite{Xiao2024}. Numerical results reveal that the dimerized model at $t_3=0$ is gapped at half filling  in the regime $|\delta J|>|\delta J_c|$ and has zero Chern number, consistent with the theoretical analysis. Whereas in the whole regime $|\delta J|\leq|\delta J_c|$, numerical results reveal gapless phase at half filling at $t_3=0$ for even $q$. At $q=2$, $\delta J_c=0$ and at $q>2$, $\delta J_c$ is finite, as indicated in Fig.\ref{phase_diagram_1} and Fig.\ref{q=4_phase_diagram}.

For the case with $t_3 \neq 0$, the sublattice symmetry is lost. Numerical analysis reveals that the phase diagrams for the general even $q$ cases are similar to the case with $q=4$ in Fig.\ref{q=4_Chern number}, i.e., at large $|\delta J|$, the system is gapped for all $\phi$, and there are generally two gap-closing points  at finite $\delta J=\pm|\delta J^\phi|$ for a given $\phi$ except for $\phi=\frac{\pi}{6}\mod \frac{\pi}{3}$ where the two gap-closing points merge to one at $\delta J=0$, as shown in Fig.\ref{phase_diagram_1} and Fig.\ref{q=4_Chern number}. At $|\delta J|\to \infty$, 
$t_3/|\delta J|\to 0$ and the system is equivalent to the square lattice Hofstadter model with $t_3=0$. For the reason, the Chern number $C=0$ at $|\delta J|\to \infty$. Without gap closing, the Chern number remains zero in the whole regime $|\delta J|>|\delta J^\phi|$, as verified by Fig.\ref{phase_diagram_1} and Fig.\ref{q=4_Chern number}.

In the middle gapped regime $|\delta J|<|\delta J^\phi|$, we can obtain a constraint on the Chern number for general even $q$ with $t_3 \neq 0$. At $\delta J=0$ and $\phi\neq \frac{\pi}{6}\mod \frac{\pi}{3}$, the model is gapped at half filling and the translation symmetry is not broken. There are generally $q$ gapped bands in the system. The Chern number of each single band satisfies the constraint \cite{ManyChern,I_Dana_1985,constraints_2026}
\begin{equation}
C_i=1\mod q, \ i=1, 2, ..., q.
\end{equation}
The total Chern number at half filling satisfies 
\begin{equation}\label{middle_chern}
C=\frac{q}{2} \mod q.
\end{equation}
This is consistent with the Chern number calculated  in Ref.\cite{Square(NNN)_Kohmoto_1990} for the triangular Hofstadter model with $\phi=0, \delta J=0$.
Without gap closing in the regime $|\delta J|<|\delta J^\phi|$, the Chern number remains the same  as for $\delta J=0$ and satisfies Eq.(\ref{middle_chern}), which is  verified in Fig.\ref{phase_diagram_1} and Fig.\ref{q=4_Chern number}.

Since the subset of inversion symmetric 1D chains exists at $\phi= \frac{n\pi}{3}, n\in Z$ for all $q$, the corresponding topological invariant ${\cal N}$ can be used to characterize the topological properties 
in the regime $\delta J<-|\delta J^\phi|$ and $\delta J>|\delta J^\phi|$, both of which have Chern number zero. For general even $q$,  ${\cal N}$
is finite at half filling at at least one of $k_1=-\frac{\pi p}{q}$ and $\pi-\frac{\pi p}{q}$ in the regime $\delta J<-|\delta J^\phi|$ for $\phi= \frac{n\pi}{3}, n\in Z$, and is zero in the regime $\delta J>|\delta J^\phi|$. Without gap closing, the whole regime $\delta J<-|\delta J^\phi|$
is then topologically non-trivial at half filling with robust edge states, whereas the whole regime $\delta J> |\delta J^\phi|$ is topologically trivial at half filling without edge states, as demonstrated in Fig.\ref{edge_states} and \ref{q=4_band}.

The results of this work may be observed in twisted bilayer TMD in the weak interaction case. 
The two valleys of the twisted TMD have opposite spin polarization, which results in opposite staggered flux $3\phi$ for electrons in the two valleys. However, from Fig.\ref{phase_diagram_1}, Fig.\ref{q=4_Chern number} and Fig.\ref{q=3_Chern_number}, 
we can see that the phase diagram remains the same if $\phi$ changes sign. The electrons in the two valleys then have the same Hall conductivity and band structure.
By tuning the displacement field between the layers, $\phi$ may change continuously and 
the Hall conductivity changes discontinuously at each phase boundary. The topologically non-trivial phase with zero Hall conductivity can be distinguished from the topologically trivial phase by its finite longitudinal conductivity  when the Fermi energy lies in the gap.

The triangular Hofstadter model with staggered flux may also be realized in cold atom systems  in optical lattices shown in \cite{Ultracold_Atomic,Cold_Atom, Ketterle2013}, or photonic crystal in \cite{Photonic_crystal}. These systems provide a variety of tunable and controllable platforms to verify the theoretical results in the triangular Hofstadter model we study in this work. 

This work focuses on the non-interacting case of the triangular Hofstadter model with staggered flux. With interaction, a variety of correlated phases may appear in the system. For example, it was proposed  that the fractional quantum Hall state may exist in the Harper-Hofstadter model with interaction~\cite{Gerster2017, Moller2019}. Besides,  the chiral spin liquid state and antiferromagnetic Mott insulator state were also proposed to exist in the triangular Hofstadter-Hubbard model in a recent work\cite{Millis2026}. Whether these or other new correlated states exist in the triangular Hofstadter model with staggered flux when interaction is included remains to be studied in the future.

\section{Summary}
To sum up, we studied the topological properties and phase diagram of the triangular Hofstadter model with staggered flux $3\phi$, which may reveal the behavior of the twist bilayer transition metal dichalcogenides in a magnetic field in a certain range of bilayer electric displacement field. This model can be mapped to a square lattice Hofstadter model with nearest neighbor plus next nearest neighbor hopping in one unique diagonal direction, the latter hopping term breaks the sublattice symmetry of the system and results in significantly different properties from the square lattice Hofstadter model with only nearest neighbor hoppings. The breaking of sublattice symmetry in the triangular Hofstadter model results in asymmetric Hofstadter spectrum in the general case except at specific staggered flux values $\phi=\frac{\pi}{6} \mod \frac{\pi}{3}$, in which case an additional symmetry ${\cal P}$ emerges and leads to symmetric Hofstadter spectrum and gapless Dirac points at $E=0$. Breaking the ${\cal P}$ symmetry by dimerization of the hopping parameters in one direction results in rich topological phases in the model.

 We then studied the dimerized triangular Hofstadter model with staggered flux and get the common feature of the system under rational external magnetic flux $\Phi_B=2\pi p/q$ as follows. For even $q$, the phase diagram of the dimerized model contains three different gapped regimes separated by two gapless phase boundaries:  one gapped regime at small dimerization with finite Chern number and two gapped regimes at large dimerization with zero Chern number. 
 For odd $q$, the system is gapless without dimerization and gapped at any finite dimerization. There are then only two  gapped regimes for odd $q$, both of which have zero Chern number. For both even and odd $q$, the two gapped regimes with zero Chern number can be further characterized by the inversion symmetry 
 of the parameterized 1D chains at $k_1=-\frac{p \pi}{q}\mod \pi$ and $\phi= \frac{n\pi}{3}, n\in Z$, which indicates that one of the two regimes is topologically non-trivial and the other is trivial. 

The results of this work may be tested in twisted bilayer TMD \cite{TMD3} with weak interaction or cold atom systems \cite{Ultracold_Atomic,Cold_Atom, Ketterle2013} in optical lattice or photonic crystals \cite{Photonic_crystal}  realized experimentally in recent years. 

\section{Acknowledgment}
We thank Jun-Wei Li for helpful discussions. This work is supported by the National Natural Science Foundation of China under Grant No. 11974166 and the Natural Science Foundation of Jiangsu Province under Grant No.BK20231398.
    
\begin{widetext}
\appendix

\section{The ${\cal P}$ symmetry at $\phi=\frac{\pi}{6}\mod \frac{\pi}{3}$}
 The triangular Hofstadter Hamiltonian with hopping $\{t_1,t_2,t_3 e^{3i\phi}\}$ along the three directions $\mathbf{a}_1, \mathbf{a}_2, \mathbf{a}_3$
in the magentic field with rational flux $\Phi_B=2\pi p/q$ is 
\begin{equation}
     H^{\phi} =-\sum\limits_{m,n} (t_1 c_{m+1,n}^{\dagger}e^{in\Phi_B}+t_2 c_{m,n+1}^{\dagger}  +t_3 e^{3i\phi} c_{m-1,n-1}^{\dagger}e^{i\Phi_{B}(1/2-n)} )c_{m,n} +h.c..
\end{equation}

 After Fourier transformation to the momentum space of the BZ, the field operators become
\begin{equation}
c_{m,n}^{\dagger} =\frac{1}{\sqrt{N}}\sum\limits_{k_1,k_2}e^{i(k_1 m+ k_2 n)}c_{k_1,k_2}^{\dagger}, 
  \  c_{m,n} =\frac{1}{\sqrt{N}}\sum\limits_{k_1,k_2}e^{-i(k_1 m+k_2 n)}c_{k_1,k_2},\ 
\end{equation}
where $k_1\in[0,2\pi),\ k_2\in [0,2\pi)$. 
The hopping terms become
\begin{equation} 
t_1\sum\limits_{m,n}c_{m+1,n}^{\dagger}c_{m,n}e^{in\Phi_B}  =t_1\sum\limits_{k_1,k_2}e^{ik_1}c_{k_1,k_2}^{\dagger}c_{k_1,k_2+\Phi_B},
\end{equation}
\begin{equation}
t_2\sum\limits_{m,n}c_{m,n+1}^{\dagger}c_{m,n}=t_2\sum\limits_{k_1,k_2}e^{ik_2}c_{k_1,k_2}^{\dagger}c_{k_1,k_2},
\end{equation}
\begin{equation}
t_3 e^{3i\phi}\sum\limits_{m,n}c_{m-1,n-1}^{\dagger}c_{m,n}e^{i\Phi_{B}(1/2-n)}= t_3e^{3i\phi}\sum\limits_{k_1,k_2}e^{-i(k_1+k_2+\Phi_{B}/2)}c_{k_1,k_2+\Phi_{B}}^{\dagger}c_{k_1,k_2}.
\end{equation}

We then fold the BZ from $k_1, k_2\in[0,2\pi)$ to
 the MBZ with $k'_1\in[0,2\pi),\ k'_2\in [0,2\pi/q)$ by the replacement 
\begin{equation}
\sum\limits_{k_1,k_2} \xrightarrow{(k_1,k_2)\to (k'_1,\ k'_2+j\Phi_{B}\ {\rm mod}\ 2\pi)}     
\sum\limits_{k'_1,k'_2}\sum\limits_{j=1}^{q},
\end{equation}
which yields
\begin{equation} 
t_1\sum\limits_{k_1,k_2}e^{ik_1}c_{k_1,k_2}^{\dagger}c_{k_1,k_2+\Phi_B}=t_1\sum\limits_{k'_1,k'_2}\sum\limits_{j=1}^{q}e^{ik'_1}c_{k'_1,k'_2+j\Phi_{B}}^{\dagger}c_{k'_1,k'_2+(j+1)\Phi_{B}},
\end{equation}
\begin{equation}
t_2\sum\limits_{k_1,k_2}e^{ik_2}c_{k_1,k_2}^{\dagger}c_{k_1,k_2}= t_2\sum\limits_{k'_1,k'_2}\sum\limits_{j=1}^{q} e^{i(k'_2+j\Phi_{B})}c_{k'_1,k'_2+j\Phi_{B}}^{\dagger}c_{k'_1,k'_2+j\Phi_{B}}, 
\end{equation}
\begin{equation}
t_3e^{3i\phi}\sum\limits_{k_1,k_2}e^{-i(k_1 +k_2 +\Phi_{B}/2)}c_{k_1,k_2+\Phi_{B}}^{\dagger}c_{k_1,k_2}=t_3 e^{3i\phi}\sum\limits_{k'_1,k'_2}\sum\limits_{j=1}^{q}e^{-i(k'_1+k'_2+(j+\frac{1}{2})\Phi_{B})}c_{k'_1,k'_2+(j+1)\Phi_{B}}^{\dagger}c_{k'_1,k'_2+j\Phi_{B}}.
\end{equation}

Introducing a $q$-component spinor 
$\Psi_{k_1,k_2}=
   (c_{k_1,k_2+\Phi_{B}}, c_{k_1,k_2+2\Phi_{B}}, \cdots, c_{k_1,k_2+q\Phi_{B}})^T$ where $(k_1, k_2)\in \rm MBZ$.
The Hamiltonian can be expressed in terms of magnetic translation operators  as
\begin{equation} 
H^{\phi}=\sum\limits_{k_1,k_2}\Psi_{k_1,k_2}^{\dagger}\mathcal{H}^{\phi}(k_1,k_2)\Psi_{k_1,k_2},
\end{equation}
with
\begin{equation}
\mathcal{H}^{\phi}(k_1,k_2)=-[t_1e^{ik_1}\mathbf{T}_1+t_2e^{ik_2}\mathbf{T}_2+t_3 e^{3i\phi} e^{-i(k_1+k_2)}\mathbf{T}_3]+h.c.,
\end{equation}
where the magnetic translation operators $\mathbf{T}_1,\mathbf{T}_2,\mathbf{T}_3$ are defined by their action on the spinor as:
\begin{equation}
\Psi_{k_1,k_2}^{\dagger}\mathbf{T}_1\Psi_{k_1,k_2}=\sum\limits_{j=1}^{q} c_{k_1,k_2+j\Phi_{B}}^{\dagger}c_{k_1,k_2+(j+1)\Phi_{B}},
\end{equation}
\begin{equation}
\Psi_{k_1,k_2}^{\dagger}\mathbf{T}_2\Psi_{k_1,k_2}=\sum\limits_{j=1}^{q} e^{i\cdot j\Phi_{B}}c_{k_1,k_2+j\Phi_{B}}^{\dagger}c_{k_1,k_2+j\Phi_{B}},
\end{equation}
\begin{equation}
\Psi_{k_1,k_2}^{\dagger}\mathbf{T}_3\Psi_{k_1,k_2}=\sum\limits_{j=1}^{q}e^{-i (j+\frac{1}{2}\Phi_{B})}c_{k_1,k_2+(j+1)\Phi_{B}}^{\dagger}c_{k_1,k_2+j\Phi_{B}}.
\end{equation}
Explicitly, in the $q$-component spinor basis, the representations of the three magnetic translation operators are
\begin{equation} \label{translation_representation}
(\mathbf{T}_1)_{m,n}=\delta_{m+1,n\ \mathrm{mod}\ q},\ (\mathbf{T}_2)_{m,n}=\delta_{m,n}e^{im\Phi_{B}}, \ (\mathbf{T}_3)_{m,n}=\delta_{m,n+1\ \mathrm{mod}\ q} e^{-i(n+\frac{1}{2})\Phi_{B}}.
\end{equation}

These magnetic translation operators satisfy the algebraic relations given in Refs.\cite{Zak,Wen1989,constraints_2026},
\begin{equation}
\mathbf{T}_1\mathbf{T}_1^{\dagger}=1,\ \mathbf{T}_2\mathbf{T}_2^{\dagger}=1,\ \mathbf{T}_3\mathbf{T}_3^{\dagger}=1,\   
\mathbf{T}_1\mathbf{T}_2=e^{i\Phi_{B}}\mathbf{T}_2\mathbf{T}_1,\ \mathbf{T}_1\mathbf{T}_2\mathbf{T}_3=e^{i\Phi_{B}/2}.
\end{equation}

 By B\'ezout's identity, there exist integers $r,s$ such that $qr+ps=1$. Using the commutation relations one finds
\begin{equation}
    \mathbf{T}_2\mathcal{H}^{\phi}(k_1,k_2)\mathbf{T}_2^{\dagger}= \mathcal{H}^{\phi}(k_1-2\pi\frac{p}{q},k_2),\  \mathbf{T}_2^{s}\mathcal{H}^{\phi}(k_1,k_2)(\mathbf{T}_2^\dagger)^s=\mathcal{H}^{\phi}(k_1-\frac{2\pi}{q},k_2).
\end{equation}

Hence the spectrum is invariant under the shift $(k_1,k_2)\to (k_1-\frac{2\pi}{q},k_2)$ and $(k_1,k_2)\to (k_1,k_2+\frac{2\pi}{q})$.

 From the matrix representation of the translation operators in Eq.(\ref{translation_representation}), one can get the commutation relationship between $\mathbf{T}_i, i=1, 2, 3$ and the ${\cal P}$ operator in the main text, as shown in Eq.(\ref{P_T_commutation}).
 From Eq.(\ref{P_T_commutation}), one can get that the Hamiltonian ${\cal H}^\phi$ satisfies the symmetry Eq.(\ref{Additional_Symmetry})  at $\phi=\frac{\pi}{6}\mod \frac{\pi}{3}$.

\section{Secular equation of the undimerized model}
In this Appendix, we calculate the secular equation of Hamiltonian 
(\ref{Hk}) of the undimerized model, i.e.,
\begin{equation}
\det([\mathcal{H}^{\phi}(k_1,k_2)]-E\mathbf{I}_{q\times q} )=0. 
\end{equation}

By calculation, the part of the determinant that depends on $k_2$ is given by  
\begin{equation}
    2(-1)^{q-1}\text{Re}\left(e^{iqk_2}\prod\limits_{j=1}^{q}B_{j}
    \right)
\end{equation}
where $B_j=-t_{2}-t_{3}e^{i(2\pi j \frac{p}{q}+\alpha)},\ j=1,2,\cdots,q$, and
 $\alpha\equiv k_1+\pi p/q-3\phi$.

From the identity \cite{Square(NNN)_Kohmoto_1990}
\begin{equation} \prod\limits_{j=1}^{q}(x^{2}-xye^{i(2\pi p/q\cdot j +\alpha)})=x^{2q}-x^{q}y^{q}e^{iq\alpha},
\end{equation}
we get 
\begin{equation}
     (-1)^{q} \prod\limits_{j=1}^{q}B_{j}
=\prod\limits_{j=1}^{q}(t_{2}+t_{3}e^{i(2\pi p/q\cdot j+\alpha)})
=t_{2}^{q}-t_{3}^{q}e^{iq(k_1-3\phi)}\cdot(-1)^{q+p}.
\end{equation}

 The part of the determinant depending on $k_2$ is then  given by  
 \begin{equation}
       2(-1)^{q-1}\text{Re}\left(e^{iqk_2}\prod\limits_{j=1}^{q}B_{j}
\right)
=-2t_2^q\cos(qk_2)+2(-1)^{p+q}t_3^q\cos(qk_1+qk_2-3q\phi).
 \end{equation}

The system possesses a dual transformation $(k_1,k_2,t_1,t_2,t_3,\phi)\to (k_2,k_1,t_2,t_1,t_3,\phi),$ under which the eigenvalue equation $\det(\mathcal{H}^{\phi}(k_1,k_2)-E\mathbf{I}_{q\times q})=0$ remains invariant. Consequently, the secular equation can be written as  
\begin{equation}
   F(E)- f(k_1,k_2)=0,\ f(k_1,k_2)=2t_1^q\cos(qk_1)+2t_2^q\cos(qk_2)-2(-1)^{p+q}t_3^q\cos(qk_1+qk_2-3q\phi),\label{fk2}
\end{equation}
where $F(E)$ is a $q$th-order polynomial of $E$ which does not depend on $(k_1,k_2)$. 

In the specific case $\phi=\frac{\pi}{6}+\frac{n\pi}{3}, n\in Z$,
\begin{equation}
f (\frac{\pi}{2},\frac{\pi}{2} )\Big|_{\phi=\frac{\pi}{6}+\frac{n\pi}{3}}
=
\begin{cases}
2(-1)^{\frac{q}{2}} \cdot (t_1^q + t_2^q + t_3^q), & q \in \rm  even, \\ 
0, &  q \in \rm  odd.
\end{cases}
\end{equation}

Since there always exists a zero mode at $(k_1,k_2)=(\frac{\pi}{2},\frac{\pi}{2})$ for any values of $t_1,t_2$ and $t_3$ when $\phi=\frac{\pi}{6}+\frac{n\pi}{3},\ n\in Z$, we get 
\begin{eqnarray}
    F(E=0)&=& \begin{cases}
    2(-1)^{q/2}(t_1^{q}+t_2^{q}+t_3^{q}), & q \in \rm  even, \\
    0, & q \in \rm  odd.
    \end{cases}
    \label{fk2}
\end{eqnarray}

\end{widetext}

\end{document}